\documentstyle[amssymb,12pt,thmsa]{article}
\input{tcilatex}
\begin{document}

\author{Mark Auslender \\
%EndAName
Department of Electrical and Computer Engineering, \\
Ben-Gurion University of the Negev\\
POB 653, Beer-Sheva 84105, Israel\\
marka@ee.bgu.ac.il}
\title{Equivalence of Two Nonequilibrium Ensembles Based on Maximum Entropy
Principle}
\date{\today }
\maketitle

\begin{abstract}
\noindent The relation between two versions of so called non-equilibrium
statistical operator method (NESOM), NESOM-1 due to Zubarev (1961) and
NESOM-2 due to Zubarev and Kalashnikov (1970), is considered. It is proved
that, once the balance equations of NESOM-2 are satisfied, those of NESOM-1
will be satisfied with the same set of the macro-parameters. The proof uses
the convexity-type inequalities, and does not involve any assumptions
additional to the rationales behind NESOM. However, converse statement
cannot be proved within this technique. An extension of the proof to overall
equivalence of the two nonequilibrium ensembles is discussed.

{\it Dedicated to the memory of Dmitrii Nikolaevich Zubarev}
\end{abstract}

\newpage {}\baselineskip 18pt{}

\bigskip \noindent {\bf {\LARGE Preface}}{}

\mathstrut \noindent The methods of non-equilibrium statistical ensembles
based on maximum entropy (MaxEnt for short) principle are nowadays well
documented. One of them, developed independently by Zubarev \cite{zubarev3}
and McLennan \cite{mclennan2} extended the Gibbs ensembles method to
non-equilibrium. Later method, due to Zubarev and Kalashnikov 
\cite[main text]{kalashnikov1}, \cite[main text]{zubarev2} made also use of
the motion quasi-invariant construction \cite{zubarev3} but applied it to
the MaxEnt distribution rather than to the entropy itself. Both developments
are referred to as Zubarev's Nonequilibrium Statistical Operator Method
(NESOM), while the notations NESOM-1 for the earlier and NESOM-2 for the
later version of NESOM, respectively, were introduced to distinct them \cite
{kalashnikov3}. Other methods, which are close to NESOM, were published in
the interim \cite{robertson}, \cite{pelyats}, \cite{kawgunt}.

When NESOM-2 became a method of choice, the expected question of its
equivalence to NESOM-1 was promptly addressed that time \cite{kalashnikov3}.
Other proofs of the equivalence, restricted to classical systems, were
published later \cite{bitensky}, \cite{tischenko2}. More later, the present
author and Kalashnikov showed that the above mentioned proofs used
assumptions, not found among the NESOM rationales, and proposed another
proof \cite{auslender2}, which seems satisfactory from the viewpoint of
rigor adopted in physics. Though available in English, Ref.\cite{auslender2}
remained un-noted - the recent review \cite{luzzi}{\it \ }refers the proofs
of Refs.\cite{kalashnikov3}, \cite{tischenko2} to as the ultimate results on
the problem. For that, and not only that, reason I decided to republish Ref.%
\cite{auslender2} on Internet. My motivation for this is better explained by
the passage below.

{\it Albeit much more applications were treated using NESOM-1 than NESOM-2},%
{\it \ no working perturbation technique beyond lowest-order approximation
(such as high-temperature or weak-interaction approximations) were developed
in the former method}.{\it \ Moreover},{\it \ because of rather involved
statistical distribution structure in NESOM-1},{\it \ such a development
seems desperate}.{\it \ On the other hand},{\it \ the statistical
distribution construction in NESOM-2 is certainly analogous to the scattered
wave function construction in the Gell-Mann and Goldberger form of
scattering theory, the MaxEnt distribution being an analog of the incoming
wave function}.{\it \ Due that analogy},{\it \ the Gell-Mann-Goldberger
integral equation based perturbation technique applies to NESOM-2 }%
\cite[main text]{kalashnikov1}. {\it So},{\it \ once the equivalence of
NESOM-1 and NESOM-2 is proved},{\it \ there is no more need to deal with
cumbersome and intractable perturbation series of NESOM-1}.{\it \ }

{\it Just within the NESOM-2 framework, on such important examples as plasma
screening and Kondo effect at non-equilibrium, partial summation of the
perturbation series was shown to be performable }\cite{fortschritte}. {\it %
However,\ this trend was not focused on since then. As a result},{\it \ in
spite of its conceptual advantages }({\it see review by Luzzi et al. }\cite
{luzzi}),{\it \ NESOM legs behind\ of regular methods }({\it such as the
Keldysh and Kadanoff-Baym ones}) {\it regarding the state-of-art respect}.%
{\it \ Nevertheless I believe that the potential of NESOM-2 }({\it contrary
to NESOM-1}){\it \ has not yet been exhausted and that a proper diagrammatic
technique development would benefit NESOM}.

The present publication is rather a `remake' than reproduction of Ref.\cite
{auslender2}. The text of the paper is essentially revised but the
references remained in the format of the original. (In Preface the citation
of Russian papers and books, even cited in the paper text, done on their
English translations, when available). Upon revision I did not aimed to make
the proof as rigorous as the modern Mathematical Physics standard requires 
\cite{simon}, since nowadays the level of rigor customary for Equilibrium
Statistical Mechanics \cite{ruelle2} still remains unreachable for most
problems of Nonequilibrium Statistical Mechanics.

\vspace{0.3in}\noindent \noindent \noindent \noindent {\bf {\LARGE %
Acknowledgment}}

\bigskip \noindent I acknowledge Vladimir Petrovich Kalashnikov for many
years of fruitful collaboration.

\baselineskip 12pt{}

\newpage {}\baselineskip 24pt{}

\section{Introduction}

NESOM is formally based on special construction, which involves the so
called quasi-equilibrium statistical operator (QESO) 
\begin{equation}
\rho _{q}\left( t,0\right) =e^{-S\left( t,0\right) }\,,\;S\left( t,0\right)
=\Phi \left( t\right) +\sum\limits_{n}F_{n}\left( t\right) P_{n}\,.
\label{qeso}
\end{equation}
Here $t$ is the time variable (which is dummy one unless dynamics enters the
play), $\left\{ P_{n}\right\} $ is a set of {\em gross variables}, that is
observables expectation values of which describe the non-equilibrium state
of interest at instant $t$. $\Phi \left( t\right) $ is the Massieur-Planck
function 
\begin{equation}
\Phi \left( t\right) =\ln \mbox{Tr}\,\left[ e^{-\sum\limits_{n}F_{n}\left(
t\right) P_{n}}\right]  \label{norm}
\end{equation}
and $F_{n}\left( t\right) $ are the macro-parameters conjugated, in
thermodynamical sense, to the gross variables averages $\left\langle
P_{n}\right\rangle _{q}^{t}$ . The first set of the equations, connecting $%
F_{n}\left( t\right) $ and $\left\langle P_{n}\right\rangle _{q}^{t}$,
follows from Eq.(\ref{norm}) 
\begin{equation}
\left\langle P_{n}\right\rangle _{q}^{t}=\mbox{Tr}\left[ \rho _{q}\left(
t,0\right) P_{n}\right] =-\frac{\partial \Phi \left( t\right) }{\partial
F_{n}\left( t\right) }\,.\,  \label{thermeq1}
\end{equation}
To arrive at the second set, define over the whole set of density matrices $%
\rho $ the information entropy functional 
\begin{equation}
{\frak E}\left[ \rho \right] =-\mbox{Tr}\left( \rho \ln \rho \right) \,,
\label{entropy}
\end{equation}
and the quasi-equilibrium entropy 
\begin{equation}
\Sigma \left( t\right) ={\frak E}\left[ \rho _{q}\right] =\left\langle
S\left( t,0\right) \right\rangle _{q}^{t}=\Phi \left( t\right)
+\sum\limits_{n}F_{n}\left( t\right) \left\langle P_{n}\right\rangle
_{q}^{t};  \label{entrfun}
\end{equation}
thus $S\left( t,0\right) $ may be called the entropy operator. Then it is
proven that 
\begin{equation}
F_{n}\left( t\right) =\frac{\partial \Sigma \left( t\right) }{\partial
\left\langle P_{n}\right\rangle _{q}^{t}}\,.  \label{thermeq2}
\end{equation}

Consider now the dependence of the operators on time. Any operator $Q$ may
explicitly depend on $t$ that is notated by $Q\left( t,0\right) $ (this
notation has already been used above for QESO and the entropy operator).
Operator $Q\left( t,0\right) $ is said to be dynamics invariant (or
integral) if it satisfies the Liouville equation 
\begin{equation}
\stackrel{\bullet }{Q}\left( t,0\right) =\frac{\partial Q(t,0)}{\partial t}+i%
{\frak L}\left( t\right) Q(t,0)=0\,,  \label{liouveq}
\end{equation}
where ${\frak L}\left( t\right) $ is the Liouville super-operator, which
action is given by the commutator (in classical case Poisson-bracket) with
the Hamiltonian. In particular, any non-equilibrium statistical operator
(NESO) satisfies Eq.(\ref{liouveq}) and QESO does not. One of the crucial
concepts in NESOM is the dynamics quasi-invariant (or quasi-integral). Given
an operator $Q\left( t,0\right) $ and a positive number $\varepsilon $,
define the dynamics quasi-integral associated with $Q\left( t,0\right) $ by 
\begin{equation}
\stackrel{\varepsilon }{\widetilde{Q\left( t,0\right) \ }}=\varepsilon
\int\limits_{-\infty }^{t}e^{\varepsilon \left( t_{0}-t\right) }Q\left(
t_{0},t_{0}-t\right) dt_{0}=Q\left( t,0\right) -\int\limits_{-\infty
}^{t}e^{\varepsilon \left( t_{0}-t\right) }\stackrel{\bullet }{Q}\left(
t_{0},t_{0}-t\right) dt_{0}\,,  \label{quasinv}
\end{equation}
where $Q(t_{1},t_{1}-t_{2})={\frak U}\left( t_{1},t_{2}\right) Q\left(
t_{1},0\right) $ and ${\frak U}\left( t_{1},t_{2}\right) $ is the dynamic
evolution super-operator. It satisfies the couple of dual to each other
equations 
\begin{equation}
\frac{\partial }{\partial t_{1}}{\frak U}\left( t_{1},t_{2}\right) ={\frak U}%
\left( t_{1},t_{2}\right) i{\frak L}\left( t_{1}\right)  \label{timevol}
\end{equation}
$\;$and 
\begin{equation}
\frac{\partial }{\partial t_{2}}{\frak U}\left( t_{1},t_{2}\right) =-i{\frak %
L}\left( t_{2}\right) {\frak U}\left( t_{1},t_{2}\right)  \label{dtimevol}
\end{equation}
with the initial conditions ${\frak U}\left( t_{2},t_{2}\right) ={\frak U}%
\left( t_{1},t_{1}\right) ={\frak I}{\cal \,}$, where and ${\frak I}$ is the
identity super-operator. For the time-independent Hamiltonians, Eqs.(\ref
{timevol}), (\ref{dtimevol}) give ${\frak U}\left( t_{1},t_{2}\right)
=e^{i\left( t_{1}-t_{2}\right) {\frak L}}$. As seen from Eqs.(\ref{liouveq})
and (\ref{quasinv}), $\stackrel{\varepsilon }{\widetilde{Q\left( t,0\right)
\ }}=Q\left( t,0\right) $ if $Q\left( t,0\right) $ is the dynamics
invariant. Moreover, $\stackrel{\varepsilon }{\widetilde{Q\left( t,0\right)
\ }}$satisfies the equation 
\begin{equation}
\frac{\partial }{\partial t}\stackrel{\varepsilon }{\widetilde{Q\left(
t,0\right) \ }}+\;i{\frak L}\left( t\right) \stackrel{\varepsilon }{%
\widetilde{Q\left( t,0\right) \ }}=\varepsilon \left[ Q\left( t,0\right) -%
\stackrel{\varepsilon }{\widetilde{Q\left( t,0\right) \ }}\right] \,
\label{quasiliouveq}
\end{equation}
approaching to Eq.(\ref{liouveq}) at $\varepsilon \downarrow 0$ that
explicates using the term `quasi-invariant'.

For a system occupying a finite domain $\Omega $ of volume $\left| \Omega
\right| $, NESOM builds from QESO so called quasi-non-equilibrium
statistical operator (QNESO) using the procedure of Eq.(\ref{quasinv}).
QNESO is denoted here by the notation $^{(\alpha )}\!\rho _{\varepsilon
}\left( t,0\right) $, where the superscript $\alpha =1,2$ points to the
NESOM version. More historically earlier one (NESOM-1) was formulated in the
papers of D.N. Zubarev \cite{zubarev1} and McLennan \cite{mclennan1} as the
generalization of canonical ensemble method to non-equilibrium. In terms of
Eq.(\ref{quasinv}) QNESO-1 is quasi-canonical distribution of the form 
\begin{equation}
^{(1)}\!\rho _{\varepsilon }\left( t,0\right) =e^{\Psi _{\varepsilon }\left(
t\right) -\stackrel{\varepsilon }{\widetilde{S\left( t,0\right) \ }}},\;\Psi
_{\varepsilon }\left( t\right) =-\ln \mbox{Tr}\left[ e^{-\stackrel{%
\varepsilon }{\widetilde{S\left( t,0\right) \ }}}\right] .  \label{neso-1}
\end{equation}
The later version (NESOM-2) was developed by D.N. Zubarev and V.P.
Kalashnikov \cite{kalashnikov1}. QNESO-2 is the dynamics quasi-integral
built from $\rho _{q}(t,0)$, that is 
\begin{equation}
^{(2)}\!\rho _{\varepsilon }\left( t,0\right) =\,\stackrel{\varepsilon }{%
\widetilde{\rho _{q}(t,0)}}\,=\,\stackrel{\varepsilon }{\widetilde{%
e^{-S(t,0)}}}.  \label{neso-2}
\end{equation}
Eq.(\ref{neso-1}) includes additional normalization to the unity trace. The
logarithm of the QNESO-1 trace normalization factor $\Psi _{\varepsilon }(t)$
plays a crucial role in the present paper. In Eq.(\ref{neso-2}) the
normalization holds automatically.

The equation for $^{(2)}\!\rho _{\varepsilon }\left( t,0\right) $ results
directly from Eq.(\ref{quasiliouveq}): 
\begin{equation}
\left[ \frac{\partial }{\partial t}+i{\frak L}\left( t\right) \right]
\!^{(2)}\!\rho _{\varepsilon }\left( t,0\right) =\varepsilon \left[ \rho
_{q}(t,0)-\,^{(2)}\!\rho _{\varepsilon }\left( t,0\right) \right] \,.
\label{eqneso-2}
\end{equation}
More involved equation for $^{(1)}\!\rho _{\varepsilon }\left( t,0\right) $
is obtained by applying to the operator exponent in Eq.(\ref{neso-1}) the
rules of time differentiation and of ${\frak L}\left( t\right) $ \cite
{zubarev2}, and then Eq.(\ref{quasiliouveq}). The result is 
\begin{eqnarray}
&&\left[ \frac{\partial }{\partial t}+i{\frak L}\left( t\right) \right]
\!\,^{(1)}\!\rho _{\varepsilon }\left( t,0\right) \,  \nonumber \\
&=&\varepsilon \,\,\hspace{-0.1in}\int\limits_{0}^{1}\,^{\left( 1\right)
}\!\rho _{\varepsilon }\left( t,0\right) e^{\lambda \stackrel{\varepsilon }{%
\widetilde{S\left( t,0\right) \ }}}\left\{ ^{\left( 1\right) }\!\Delta
_{\varepsilon }^{t}\left[ \stackrel{\varepsilon }{\widetilde{S\left(
t,0\right) \ }}-\;S\left( t,0\right) \right] \right\} e^{-\lambda \stackrel{%
\varepsilon }{\widetilde{S\left( t,0\right) \ }}}d\lambda \,\,,\;
\label{eqneso-1}
\end{eqnarray}
where 
\begin{equation}
^{(\alpha )}\!\Delta _{\varepsilon }^{t}A\triangleq A-\,^{(\alpha
)}\!\left\langle A\right\rangle _{\varepsilon }^{t}  \label{delta}
\end{equation}
by the definition and 
\begin{equation}
^{(\alpha )}\!\left\langle A\right\rangle _{\varepsilon }^{t}\triangleq %
\mbox{Tr}\left[ ^{(\alpha )}\!\rho _{\varepsilon }\left( t,0\right) A\right]
.  \label{quasiav}
\end{equation}
The source-like terms emerged in the rhs of Eqs.(\ref{eqneso-1}), (\ref
{eqneso-2}) break time-reversal invariance of the Liouville equation.
Non-equilibrium may be described using $\,$expectation values of the type
given by Eq.(\ref{quasiav}) while $\varepsilon $ tends to zero. But, due the
Bogoljubov's idea of quasi-averages, any invariance-breaking perturbation is
switched off only {\em after performing the thermodynamic limit} (TL). So
the quasi-averages, which would sustain the broken symmetry of time arrow
towards future, may be defined by 
\begin{equation}
^{(\alpha )}\!\left\langle A\right\rangle ^{t}=\lim\limits_{\varepsilon
\downarrow 0}\,\lim\limits_{\left| \Omega \right| \rightarrow \infty
}{}^{(\alpha )}\!\left\langle A\right\rangle _{\varepsilon }^{t}\,.
\label{limquasiav}
\end{equation}
Of course, Eq.(\ref{limquasiav}) makes any sense if $A$ is an intensive
observable (like density of an extensive observable) that may be assumed
without any loss of generality. As was argued by D.N. Zubarev \cite{zubarev2}%
, such a subordination of the $\varepsilon $ limit ($\varepsilon $L) and TL
is crucial for the correct simulation of irreversibility.\footnote{%
TL and $\varepsilon $L may be performed simultaneously, but anyhow $%
\varepsilon \left| \Omega \right| \rightarrow \infty $ \cite{zubarev1}, 
\cite[Preface]{zubarev3}}

Eqs.(\ref{thermeq1}) and (\ref{thermeq2}) by no means define the
macro-parameters. The ultimate rationale of NESOM, which facilitates a
closed description of non-equilibrium, is the balance equations (BE). BE may
be presented in different equivalent forms. The basic one is the
self-consistency form that reads 
\begin{equation}
^{(\alpha )}\!\left\langle P_{n}\right\rangle ^{t}=\left\langle
P_{n}\right\rangle ^{t}\triangleq \lim_{\left| \Omega \right| \rightarrow
\infty }\left\langle P_{n}\right\rangle _{q}^{t}\,,  \label{baleq}
\end{equation}
but actually the differential form of BE 
\begin{equation}
\frac{\partial }{\partial t}\left\langle P_{n}\right\rangle ^{t}={}^{(\alpha
)}\!\left\langle \stackrel{\bullet }{P_{n}}\left( t,0\right) \right\rangle
^{t}  \label{difbaleq}
\end{equation}
is exploited in the practice. Slightly different route consists of using the
`pre-limit' BE 
\begin{equation}
^{(\alpha )}\!\left\langle P_{n}\right\rangle _{\varepsilon
}^{t}=\left\langle P_{n}\right\rangle _{q}^{t}\,  \label{baleqf}
\end{equation}
and 
\begin{equation}
\frac{\partial }{\partial t}\left\langle P_{n}\right\rangle
_{q}^{t}={}^{(\alpha )}\!\left\langle \stackrel{\bullet }{P_{n}}\left(
t,0\right) \right\rangle _{\varepsilon }^{t}\,,  \label{difbaleqf}
\end{equation}
at yet finite but large $\left| \Omega \right| $ and small $\varepsilon $,
and then performing TL and $\varepsilon $L in due order (see D.N. Zubarev's
review in Ref.\cite{zubarev2}). For $\alpha =2$, Eq.(\ref{difbaleq}) and Eq.(%
\ref{difbaleqf}) exactly follow from Eq.(\ref{baleq}) and Eq.(\ref{baleqf}),
respectively, (see proof in e.g. \cite{zubarev2}) while for $\alpha =1$ the
question of relation between the two forms of BE is not so simple. A
discussion on this issue is presented in Appendix I.

As seen, BE are complicated (non-local) functional equations for defining $%
F_{n}\left( t\right) $. NESOM asserts that solutions to BE, $^{(\alpha
)}\!F_{n}\left( t\right) $, do exist and finds them in physical
approximations. In the first order with respect to the entropy production,
QNESO-1 and QNESO-2 are the same \cite{zubarev2}. Thus, NESOM-1 and NESOM-2
are equivalent in that approximation. This concerns both the coincidence of
the BE\ solutions 
\begin{equation}
^{(1)}\!F_{n}\left( t\right) =\,^{(2)}\!F_{n}\left( t\right)
\label{minequiv}
\end{equation}
and of the averages over the two NESOM ensembles 
\begin{equation}
{}^{(1)}\!\left\langle A\right\rangle ^{t}=\,^{(2)}\!\left\langle
A\right\rangle ^{t}\,  \label{maxequiv}
\end{equation}
at least for some class of intensive observables. While Eq.(\ref{maxequiv})
holds, Eq.(\ref{minequiv}) holds automatically, but not vice versa. Quite
naturally, the general equivalence of NESOM ensembles is in question.
Authors of Ref.\cite{kalashnikov2} discussed a condition for Eq.(\ref
{minequiv}) to hold. In Refs.\cite{tischenko1} and \cite[Preface]{bitensky}
proofs of Eq.(\ref{maxequiv}) were suggested. As shown in Appendix II, these
proofs essentially used assumptions which, in fact, are not found among the
basics behind NESOM. Moreover, it is shown that the condition of Ref.\cite
{tischenko1} and `asymptotic normalization' condition of Ref.\cite[Preface]
{bitensky} can not be satisfied.

In this paper a proof of Eq.(\ref{minequiv}) is presented. This proof is
based only on the NESOM rationales and on a number of assumptions, which
seem to be tacitly adopted in NESOM and in other methods of reduced
description of non-equilibrium (see e.g. \cite[Preface]{robertson} - 
\cite[Preface]{kawgunt}).

\section{Pierls-Bogoljubov Inequality}

The main tool in this paper is the Pierls-Bogoljubov inequality \cite
{ruelle1} (also \cite[Preface]{ruelle2}). This inequality is used in the
two-sided form: 
\begin{equation}
\frac{\mbox{Tr}\left[ e^{-A}\left( A-B\right) \right] }{\mbox{Tr}\left(
e^{-A}\right) }\leq \ln \mbox{Tr}\left( e^{-B}\right) -\ln \mbox{Tr}\left(
e^{-A}\right) \leq \frac{\mbox{Tr}\left[ e^{-B}\left( A-B\right) \right] }{%
\mbox{Tr}\left( e^{-B}\right) }\,,  \label{PBineq}
\end{equation}
the equality being possible only at $A=B$. To begin with, put $A=-\ln
\,\left[ ^{\left( \alpha \right) }\!\rho _{\varepsilon }\left( t,0\right)
\right] $ $\,$and $B=S\left( t,0\right) $. On account of the trace
normalisation, the left side of Eq.(\ref{PBineq}) leads to the inequality 
\begin{equation}
^{\left( \alpha \right) }\!\left\langle S\left( t,0\right) \right\rangle
_{\varepsilon }^{t}>{\frak E}\left[ ^{\left( \alpha \right) }\!\rho
_{\varepsilon }\right] \triangleq \,\,^{\left( \alpha \right) }{\frak S}%
_{\varepsilon }\left( t\right) \,  \label{entrineq}
\end{equation}
that, upon satisfying BE in the form of Eq.(\ref{baleqf}), transforms to 
\begin{equation}
\Sigma \left( t\right) >\,^{\left( \alpha \right) }{\frak S}_{\varepsilon
}\left( t\right) \,,  \label{maxentp}
\end{equation}
where the lhs also becomes dependent on $\varepsilon $ (and generally on $%
\alpha $). This inequality expresses the {\em MaxEnt principle} at finite $%
\left| \Omega \right| $ and $\varepsilon $. The function $^{\left( \alpha
\right) }\!{\frak S}_{\varepsilon }\left( t\right) $ may be called
quasi-Gibbs entropy. Post TL and $\varepsilon $L form of the MaxEnt
principle, the only feasible for the NESOM's route with Eq.(\ref{baleq}),
holds for the specific entropies 
\begin{equation}
\sigma \left( t\right) \,=\,^{\left( \alpha \right) }\!\sigma \left(
t\right) =\lim_{\varepsilon \downarrow 0}\,^{\left( \alpha \right) }\!\sigma
_{\varepsilon }\left( t\right) >\lim_{\varepsilon \downarrow 0}\,^{\left(
\alpha \right) }\!{\frak s}_{\varepsilon }\left( t\right) =\,^{\left( \alpha
\right) }\!{\frak s}\left( t\right) ,  \label{maxspentp}
\end{equation}
where 
\begin{equation}
\sigma \left( t\right) =\lim_{\left| \Omega \right| \rightarrow \infty }%
\frac{\Sigma \left( t\right) }{\left| \Omega \right| },\;\,^{\left( \alpha
\right) }\!\sigma _{\varepsilon }\left( t\right) \,=\lim_{\left| \Omega
\right| \rightarrow \infty }\frac{^{\left( \alpha \right) }\!\left\langle
S\left( t,0\right) \right\rangle _{\varepsilon }^{t}}{\left| \Omega \right| }
\label{spneqentr}
\end{equation}
are the specific quasi-equilibrium and pre-$\varepsilon $L non-equilibrium
entropies, and 
\begin{equation}
\,^{\left( \alpha \right) }\!{\frak s}_{\varepsilon }\left( t\right)
=\lim_{\left| \Omega \right| \rightarrow \infty }\frac{^{\left( \alpha
\right) }{\frak S}_{\varepsilon }\left( t\right) }{\left| \Omega \right| }.
\label{spqGentr}
\end{equation}
is the specific pre-$\varepsilon $L quasi-Gibbs entropy. It is assumed that
all limits in Eqs.(\ref{maxspentp}) - (\ref{spqGentr}) exist and BE are
satisfied (see next section).

\section{Handling TL and $\varepsilon $L}

TL and $\varepsilon $L are inevitable ingredients of NESOM. In order to
accurately resolve TL issue (but, at the same time, do not enter its
subtleties \cite{ruelle1}, \cite[Preface]{ruelle2}) it is expedient to
introduce several relevant axioms, which are necessary for the NESOM
formulation to make any sense and seem quite satisfactory from the viewpoint
of rigor adopted by physicists. As to $\varepsilon $L, it is treated here in
a simplified manner (for a mathematically correct treatment, see 
\cite[soon on Web]{auslender1}).

\begin{axiom}
\noindent The index $n$ is discrete at finite $\left| \Omega \right| $ and
continuous at $\left| \Omega \right| \rightarrow \infty $, while there
exists a measure $\mu _{n}$ such that 
\begin{equation}
\lim_{\left| \Omega \right| \rightarrow \infty }\frac{1}{\left| \Omega
\right| }\sum_{n}g_{n}=\int g_{n}d\mu _{n}  \label{sumnext}
\end{equation}
for every bounded function $g_{n}$. The $F_{n}\left( t\right) $ by $%
\left\langle P_{n}\right\rangle ^{t}$ and by $\lim\limits_{\left| \Omega
\right| \rightarrow \infty }{}^{\left( \alpha \right) }\!\left\langle
P_{n}\right\rangle _{\varepsilon }^{t}$ products are $\mu $-integrable.%
\footnote{%
Although in $n$ there may be a continous species before and remain a
discrete species after TL, this axiom may be assumed without any loss of
generality.}
\end{axiom}

\begin{axiom}
\noindent $\Phi \left( t\right) $ is an extensive quasi-thermodynamic
potential, in the sense that there exists the limit 
\begin{equation}
\phi \left( t\right) =\lim_{\left| \Omega \right| \rightarrow \infty }\frac{%
\Phi \left( t\right) }{\left| \Omega \right| }\,,  \label{mas-plpot}
\end{equation}
\end{axiom}

\noindent which is an analog of the equilibrium specific grand-canonical
potential \cite{isihara}. The above axioms assure existence of the functions
defined in Eqs.(\ref{maxspentp}), (\ref{spneqentr}). Firstly $\sigma \left(
t\right) $, being an analog of the equilibrium specific entropy \cite
{isihara}, is given by 
\begin{equation}
\sigma \left( t\right) =\phi \left( t\right) +\int F_{n}\left( t\right)
\left\langle P_{n}\right\rangle ^{t}d\mu _{n}\,.  \label{spqeentr}
\end{equation}
Post TL form of Eqs.(\ref{thermeq1}) and (\ref{thermeq2}) express $%
\left\langle P_{n}\right\rangle ^{t}$ and $F_{n}\left( t\right) $ by {\em %
variational} {\em derivation} of $\phi \left( t\right) $ and $\sigma \left(
t\right) $ in $F_{n}\left( t\right) $ and $\left\langle P_{n}\right\rangle
^{t}$, respectively (keeping $t$ frozen), so that $\phi \left( t\right) $
and $\sigma \left( t\right) $ are connected to each other via Legendre
transformation, e.g. 
\begin{equation}
\phi \left( t\right) =\sigma \left( t\right) -\int \left\langle
P_{n}\right\rangle ^{t}\frac{\delta \sigma \left( t\right) }{\delta
\left\langle P_{n}\right\rangle ^{t}}d\mu _{n}\,.  \label{phi-sigma}
\end{equation}
Lastly $^{\left( \alpha \right) }\!\sigma _{\varepsilon }\left( t\right) $
and $^{\left( \alpha \right) }\!\sigma \left( t\right) $ are given by 
\begin{equation}
^{\left( \alpha \right) }\sigma _{\varepsilon }\left( t\right) =\phi \left(
t\right) +\int F_{n}\left( t\right) \lim\limits_{\left| \Omega \right|
\rightarrow \infty }\,^{\left( \alpha \right) }\!\left\langle
P_{n}\right\rangle _{\varepsilon }^{t}d\mu _{n}  \label{spneentr}
\end{equation}
and 
\begin{equation}
^{\left( \alpha \right) }\sigma \left( t\right) =\phi \left( t\right) +\int
F_{n}\left( t\right) \,^{\left( \alpha \right) }\!\left\langle
P_{n}\right\rangle ^{t}d\mu _{n}\,,  \label{Lspneentr}
\end{equation}
respectively.

\begin{axiom}
$\lim\limits_{\left| \Omega \right| \rightarrow \infty }\left\langle
P_{n}\left( t_{0},t_{0}-t_{1}\right) \right\rangle _{q}^{t_{1}}$ exists and 
{\em (}by physical reasons{\em )} is continuous bounded function of time
arguments.\footnote{%
E.g. for bounded (in the Hilbert space of the system) operators $P_{n}$,
which norms $\left\| P_{n}\right\| $ are independent of $V_{\Omega }$, the
absolute value of this function is bounded by $\left\| P_{n}\right\| $.} The
product $F_{n}\left( t_{0}\right) $ by $\lim\limits_{\left| \Omega \right|
\rightarrow \infty }\left\langle P_{n}\left( t_{0},t_{0}-t_{1}\right)
\right\rangle _{q}^{t_{1}}$ is $\mu $-integrable function of $n$.
\end{axiom}

\noindent The following corollary ensues from these axioms and the
rationales of NESOM.

\begin{corollary}
$\lim\limits_{\left| \Omega \right| \rightarrow \infty }\,^{\left( 2\right)
}\!\left\langle P_{n}\left( t_{0},t_{0}-t\right) \right\rangle _{\varepsilon
}^{t}$ exists and the product $F_{n}\left( t_{0}\right) \,$by $^{\left(
2\right) }\!\left\langle P_{n}\left( t_{0},t_{0}-t\right) \right\rangle
_{\varepsilon }^{t}$ is $\mu $-integrable function. In addition 
\begin{equation}
^{\left( 2\right) }\!\left\langle P_{n}\left( t_{0},t_{0}-t\right)
\right\rangle ^{t}=\,^{\left( 2\right) }\!\left\langle P_{n}\right\rangle
^{t_{0}}\,.  \label{propneso-2}
\end{equation}
\end{corollary}

%TCIMACRO{\TeXButton{Proof}{\proof}}
%BeginExpansion
\proof%
%EndExpansion
By the definition of Eq.(\ref{quasiav}) and the evolution super-operator
property 
\begin{equation}
^{\left( 2\right) }\!\left\langle P_{n}\left( t_{0},t_{0}-t\right)
\right\rangle _{\varepsilon }^{t}=\mbox{Tr}\left\{ P_{n}\left[ {\frak U}%
\left( t,t_{0}\right) \,^{(2)}\!\rho _{\varepsilon }\left( t,0\right)
\right] \right\} .  \label{interim1}
\end{equation}
The action of the super-operator $e^{\varepsilon \left( t-t_{0}\right) }%
{\frak U}\left( t,t_{0}\right) $ on both sides of Eq.(\ref{eqneso-2}) and
the integration of the resulting equation from $t_{0}$ to $t$ leads to 
\[
e^{\varepsilon \left( t-t_{0}\right) }{\frak U}\left( t,t_{0}\right)
\,^{(2)}\!\rho _{\varepsilon }\left( t,0\right) =\,^{(2)}\!\rho
_{\varepsilon }\left( t_{0},0\right) +\varepsilon
\int\limits_{t_{0}}^{t}e^{\varepsilon \left( t_{1}-t_{0}\right) }{\frak U}%
\left( t_{1},t_{0}\right) \rho _{q}\left( t_{1},0\right) dt_{1}\,. 
\]
The use of this operator identity and Eq.(\ref{interim1}) gives the
following functional identity 
\begin{equation}
^{\left( 2\right) }\!\left\langle P_{n}\left( t_{0},t_{0}-t\right)
\right\rangle _{\varepsilon }^{t}=\,^{\left( 2\right) }\!\left\langle
P_{n}\right\rangle _{\varepsilon }^{t_{0}}e^{\varepsilon \left(
t_{0}-t\right) }+\varepsilon \int\limits_{t_{0}}^{t}\left\langle P_{n}\left(
t_{0},t_{0}-t_{1}\right) \right\rangle _{q}^{t_{1}}e^{\varepsilon \left(
t_{1}-t\right) }dt_{1},  \label{interim2}
\end{equation}
from which the lemma statement follows, including Eq.(\ref{propneso-2}).
Note that performing $\varepsilon $L in Eqs.(\ref{spneentr}) and (\ref
{interim2}) is quite rigorous within the NESOM rationales.%
%TCIMACRO{\TeXButton{End Proof}{\endproof}}
%BeginExpansion
\endproof%
%EndExpansion

Though assuming existence of $\,\!^{\left( \alpha \right) }{\frak s}%
_{\varepsilon }\left( t\right) $ is physically reasonable, nothing assures
this formally. The statements below connect that question for $\alpha =1$ to
the behavior of $\Psi _{\varepsilon }\left( t\right) $ in TL.

\begin{lemma}
{}{}{\em (a)} Existence of \/$^{\left( 1\right) }{\frak s}_{\varepsilon
}\left( t\right) $ is equivalent to extensivity of $\Psi _{\varepsilon
}\left( t\right) $, that is to existence of 
\begin{equation}
\psi _{\varepsilon }\left( t\right) =\lim_{\left| \Omega \right| \rightarrow
\infty }\frac{\Psi _{\varepsilon }\left( t\right) }{\left| \Omega \right| }%
\,.  \label{TLpsi}
\end{equation}
{\em (b)} For both $\psi _{\varepsilon }\left( t\right) $ and $^{\left(
1\right) }{\frak s}_{\varepsilon }\left( t\right) $ to exist it is
sufficient that $\lim\limits_{\left| \Omega \right| \rightarrow \infty }$ $%
^{\left( 1\right) }\!\left\langle P_{n}\left( t_{0},t_{0}-t\right)
\right\rangle _{\varepsilon }^{t}$ exist and the $F_{n}\left( t_{0}\right) $
by $\lim\limits_{\left| \Omega \right| \rightarrow \infty }$ $^{\left(
1\right) }\!\left\langle P_{n}\left( t_{0},t_{0}-t\right) \right\rangle
_{\varepsilon }^{t}$ products are $\mu $-integrable.
\end{lemma}

%TCIMACRO{\TeXButton{Proof}{\proof} }
%BeginExpansion
\proof%
%EndExpansion
Differentiation of the definition of $\Psi _{\varepsilon }\left( t\right) $
in Eq.(\ref{neso-1}) gives the relation 
\begin{equation}
\frac{\partial }{\partial t}\Psi _{\varepsilon }\left( t\right) =\varepsilon
\!\,\left[ \,^{\left( 1\right) }\!\left\langle \;S\left( t,0\right)
\right\rangle _{\varepsilon }^{t}\,-\,^{\left( 1\right) }\!\left\langle 
\stackrel{\varepsilon }{\widetilde{S\left( t,0\right) \ }}\right\rangle
_{\varepsilon }^{t}\right] \,\!.  \label{Psider}
\end{equation}
By the definition of Eq.(\ref{entrineq}) and Eq.(\ref{Psider}) 
\begin{equation}
^{\left( 1\right) }{\frak S}_{\varepsilon }\left( t\right) =\,-\Psi
_{\varepsilon }\left( t\right) +\,^{\left( 1\right) }\!\left\langle 
\stackrel{\varepsilon }{\widetilde{S\left( t,0\right) \ }}\right\rangle
_{\varepsilon }^{t}\,=\,^{\left( 1\right) }\!\!\left\langle \;S\left(
t,0\right) \right\rangle _{\varepsilon }^{t}\,-\left( 1+\varepsilon ^{-1}%
\frac{\partial }{\partial t}\right) \Psi _{\varepsilon }\left( t\right) .
\label{qGentr-Psi}
\end{equation}
This equation may be integrated to give 
\begin{equation}
\Psi _{\varepsilon }\left( t\right) =\varepsilon \int\limits_{-\infty
}^{t}\,e^{\varepsilon \left( t_{0}-t\right) }\left[ ^{\left( 1\right)
}\!\left\langle S\left( t_{0},0\right) \right\rangle _{\varepsilon
}^{t_{0}}-\,^{\left( 1\right) }{\frak S}_{\varepsilon }\left( t_{0}\right)
\right] dt_{0}\,.  \label{Psi-qGentr}
\end{equation}
Due to Eqs.(\ref{entrineq}), (\ref{Psi-qGentr}) $\Psi _{\varepsilon }\left(
t\right) >0$, the fact that will further be proved in other way. Eqs.(\ref
{qGentr-Psi}) and (\ref{Psi-qGentr}), due to extensivity of $^{\left(
1\right) }\!\left\langle S\left( t,0\right) \right\rangle _{\varepsilon
}^{t} $ expressed by Eq.(\ref{spneentr}), show that existence of $\psi
_{\varepsilon }\left( t\right) $ does provide that of $^{\left( 1\right) }%
{\frak s}_{\varepsilon }\left( t\right) $ and vice versa. Eq.(\ref{quasinv})
and Axiom 1 give 
\begin{equation}
\lim_{\left| \Omega \right| \rightarrow \infty }\frac{1}{\left| \Omega
\right| }^{\left( 1\right) }\!\left\langle \stackrel{\varepsilon }{%
\widetilde{S\left( t,0\right) \ }}\right\rangle _{\varepsilon
}^{t}=\varepsilon \int\limits_{-\infty }^{t}\,e^{\varepsilon \left(
t_{0}-t\right) }\int F_{n}\left( t_{0}\right) \,\,^{\left( 1\right)
}\left\langle P_{n}\left( t_{0},t_{0}-t\right) \right\rangle _{\varepsilon
}^{t}d\mu _{n}dt_{0},  \label{spStild}
\end{equation}
provided that the condition of (b) holds. As seen from the above
consideration, Eq.(\ref{spStild}) assures existence of both $\psi
_{\varepsilon }\left( t\right) $ and $^{\left( 1\right) }{\frak s}%
_{\varepsilon }\left( t\right) $. On physical level of rigor, the condition
of (b) may be adopted as being also necessary for (a). It is worth
emphasizing that an attempt to prove the analog of Corollary 1 for $^{\left(
1\right) }\!\left\langle P_{n}\left( t_{0},t_{0}-t\right) \right\rangle
_{\varepsilon }^{t}$, making use of Axioms 1-3 alone, fails.%
%TCIMACRO{\TeXButton{End Proof}{\endproof}}
%BeginExpansion
\endproof%
%EndExpansion

In TL Eq.(\ref{Psi-qGentr}) and the first equality in Eq.(\ref{qGentr-Psi})
transform to 
\begin{equation}
\psi _{\varepsilon }\left( t\right) =\varepsilon \int\limits_{-\infty
}^{t}\,e^{\varepsilon \left( t_{0}-t\right) }\left[ ^{\left( 1\right)
}\sigma _{\varepsilon }\left( t_{0}\right) -\,^{\left( 1\right) }{\frak s}%
_{\varepsilon }\left( t_{0}\right) \right] dt_{0}  \label{TLpsi-spqGentr}
\end{equation}
and 
\begin{equation}
\,^{\left( 1\right) }\!{\frak s}_{\varepsilon }\left( t\right)
=\,\varepsilon \int\limits_{-\infty }^{t}\,e^{\varepsilon \left(
t_{0}-t\right) }\sigma \left( t_{0}\right) dt_{0}-\,\psi _{\varepsilon
}\left( t\right) +\int\limits_{-\infty }^{t}\int F_{n}\left( t_{0}\right) 
\frac{\delta \psi _{\varepsilon }\left( t\right) }{\delta F_{n}\left(
t_{0}\right) }d\mu _{n}dt_{0}\,,  \label{Legtrans}
\end{equation}
respectively. Unfortunately nothing may be inferred from Eqs.(\ref
{TLpsi-spqGentr}), (\ref{Legtrans}) on existence of $^{\left( 1\right) }%
{\frak s}\left( t\right) $ and 
\begin{equation}
\psi \left( t\right) =\lim_{\varepsilon \downarrow 0}\psi _{\varepsilon
}\left( t\right) ,  \label{eLpsi}
\end{equation}
except the fact that these quantities, if exist, should be independent of
time: $\psi \left( t\right) =\psi $ and $^{\left( 1\right) }{\frak s}\left(
t\right) =\,^{\left( 1\right) }{\frak s}$.

\section{Equivalence of Two NESOM Ensembles}

\subsection{Generating Functional in NESOM-1}

Let ${\frak f}=\left\{ f_{n}\right\} $ is a set of functions independent of
time, satisfying the same requirements as $F_{n}\left( t\right) $. Define
upon ${\frak f}$ the functional 
\begin{equation}
\Psi _{\varepsilon }\left( t;{\frak f}\right) =-\ln \mbox{Tr}\left[
e^{-S_{\varepsilon }\left( t,{\frak f}\right) }\right] ,  \label{Psigf}
\end{equation}
where 
\begin{equation}
S_{\varepsilon }\left( t;{\frak f}\right) \,=\;\stackrel{\varepsilon }{%
\widetilde{S\left( t,0\right) \ }}+\sum_{n}f_{n}P_{n}\,.  \label{entrpert}
\end{equation}
Formally, $\Psi _{\varepsilon }\left( t;{\frak f}\right) $ is the logarithm
of trace normalisation factor for the statistical distribution with an
`entropy' defined by Eq.(\ref{Psigf}). While the use made of the last
equality in Eq.(\ref{quasinv}), this `entropy' is presented as 
\[
S_{\varepsilon }\left( t;{\frak f}\right) \,=S\left( t,0\right)
+\sum_{n}f_{n}P_{n}\,-\int\limits_{-\infty }^{t}e^{\varepsilon \left(
t_{0}-t\right) }\stackrel{\bullet }{S}\left( t_{0},t_{0}-t\right) dt_{0}\, 
\]
that may be thought as an artificial perturbation, by the shifts $%
F_{n}\left( t\right) \rightarrow F_{n}\left( t\right) +f_{n}$, concerning
the QESO entropy part with {\em no effect on the entropy production}. The
distribution under consideration is auxiliary one, which tends to `shifted'
QESO, while zeroing the entropy production. The Taylor series for $\Psi
_{\varepsilon }\left( t;{\frak f}\right) $ has the form 
\begin{equation}
\Psi _{\varepsilon }\left( t;{\frak f}\right) =\Psi _{\varepsilon }\left(
t\right) +\sum_{l=1}^{\infty }\frac{1}{l!}\sum_{n_{1}...n_{l}}K_{\varepsilon
}^{\left( l\right) }\left( t;n_{1},...,n_{l}\right) \prod_{i=1}^{l}f_{n_{i}}
\label{PsigfTs}
\end{equation}
where $\Psi _{\varepsilon }\left( t\right) $ was defined earlier and 
\begin{equation}
K_{\varepsilon }^{\left( l\right) }\left( t;n_{1},...,n_{l}\right) =\left. 
\frac{\partial ^{l}\Psi _{\varepsilon }\left( t;{\frak f}\right) }{\partial
f_{n_{1}}...\partial f_{n_{l}}}\right| _{{\frak f}=0},\,l\geq 1.
\label{lthderPsi}
\end{equation}
For particular $l=1$, 
\begin{equation}
K_{\varepsilon }^{\left( 1\right) }\left( t;n\right) =\,^{\left( 1\right)
}\left\langle P_{n}\right\rangle _{\varepsilon }^{t}\,.  \label{Pavneso-1}
\end{equation}
Higher derivatives of $\Psi _{\varepsilon }\left( t;{\frak f}\right) $ at $%
{\frak f}=0$ are connected with the cumulant correlators of gross variables
over QNESO-1 
\begin{equation}
K_{\varepsilon }^{\left( l\right) }\left( t;n_{1},...,n_{l}\right) =\left(
-1\right) ^{l-1}\,^{\left( 1\right) }\left( P_{n_{1}}\cdot ...\cdot
P_{n_{l}}\right) _{\varepsilon }^{t}\,,  \label{Ksupl}
\end{equation}
where 
\begin{equation}
\,^{\left( 1\right) }\left( A_{_{1}}\cdot ...\cdot A_{l-1}\cdot A_{l}\right)
_{\varepsilon }^{t}\triangleq
\int\limits_{0}^{1}...\int\limits_{0}^{1}\,^{\left( 1\right) }\left\langle
\left[ A_{_{1}}\left( \lambda _{1}\right) _{\varepsilon
}^{t}\,...\,A_{l-1}\left( \lambda _{l-1}\right) _{\varepsilon }^{t}\right] _{%
\text{irr}}^{>}A_{l}\right\rangle _{\varepsilon
}^{t}\prod_{i=1}^{l-1}d\lambda _{i}\,,  \label{lthCum}
\end{equation}
with 
\begin{equation}
A\left( \lambda \right) _{\varepsilon }^{t}=e^{\lambda \stackrel{\varepsilon 
}{\widetilde{S\left( t,0\right) \ }}}A\,e^{-\lambda \stackrel{\varepsilon }{%
\widetilde{S\left( t,0\right) \ }}},  \label{Aolam}
\end{equation}
and the `irreducible' ordered products calculated via the following
recursive relations 
\[
\left[ A_{_{1}}\left( \lambda _{1}\right) _{\varepsilon
}^{t}\,...\,A_{m}\left( \lambda _{m}\right) _{\varepsilon }^{t}\right] _{%
\text{irr}}^{>}=\left[ A_{_{1}}\left( \lambda _{1}\right) _{\varepsilon
}^{t}\,...\,A_{m}\left( \lambda _{m}\right) _{\varepsilon }^{t}\right]
^{>}-\,^{\left( 1\right) }\left\langle \left[ A_{_{1}}\left( \lambda
_{1}\right) _{\varepsilon }^{t}\,...\,A_{m}\left( \lambda _{m}\right)
_{\varepsilon }^{t}\right] ^{>}\right\rangle _{\varepsilon }^{t} 
\]
\begin{equation}
-\sum_{k=1}^{m-1}\sum_{\pi _{k}^{m}}\left[ A_{i_{1}}\left( \lambda
_{i_{1}}\right) _{\varepsilon }^{t}\,...\,A_{i_{k}}\left( \lambda
_{i_{k}}\right) _{\varepsilon }^{t}\right] _{\text{irr}}^{>}\,^{\left(
1\right) }\left\langle \left[ A_{i_{k+1}}\left( \lambda _{i_{k+1}}\right)
_{\varepsilon }^{t}\,...\,A_{i_{m}}\left( \lambda _{i_{m}}\right)
_{\varepsilon }^{t}\right] ^{>}\right\rangle _{\varepsilon }^{t}\,.
\label{irrprod}
\end{equation}
In these relation $\pi _{k}^{m}$ are all permutations of the type 
\begin{equation}
\pi _{k}^{m}=\left( 
\begin{array}{l}
i_{1}...i_{k}\,i_{k+1}...i_{m} \\ 
1....k\,k+1...m
\end{array}
\right) ;\;i_{1}<...<i_{k},\,i_{k+1}<...<i_{m}  \label{permut}
\end{equation}
and $\left[ B_{_{1}}\left( \lambda _{1}\right) ...\,B_{p}\left( \lambda
_{p}\right) \right] ^{>}$ is the Dyson-ordered product, in which the factors
with larger $\lambda $'s are stood to the left. Note that $^{\left( 1\right)
}\left( A\cdot B\right) _{\varepsilon }^{t}$ is a non-equilibrium analog of
the Kubo-Duhamel correlator \cite{ruelle1} (also \cite[Preface]{ruelle2}).
In spite of apparent assymetry relative to $A_{l}$ of $\,^{\left( 1\right)
}\left( A_{_{1}}\cdot ...\cdot A_{l-1}\cdot A_{l}\right) _{\varepsilon }^{t}$%
, as defined by Eq.(\ref{lthCum}), this correlator appears to be symmetric
with respect to all permutations of its operator constituents.

Consider the limit 
\begin{equation}
\psi _{\varepsilon }\left( t;{\frak f}\right) =\lim_{\left| \Omega \right|
\rightarrow \infty }\frac{\Psi _{\varepsilon }\left( t;{\frak f}\right) }{%
\left| \Omega \right| }  \label{TLpsigf}
\end{equation}
Term-by-term analysis of the series in Eq.(\ref{PsigfTs}) based upon Axiom
1, Eqs.(\ref{Pavneso-1}) and (\ref{Ksupl}) leads to the conclusion that $%
\psi _{\varepsilon }\left( t;{\frak f}\right) $ would be represented by {\em %
functional} Taylor series 
\begin{equation}
\psi _{\varepsilon }\left( t;{\frak f}\right) =\psi _{\varepsilon }\left(
t\right) +\sum_{l=1}^{\infty }\frac{1}{l!}\int ...\int \varkappa
_{\varepsilon }^{\left( l\right) }\left( t;n_{1},...,n_{l}\right)
\prod_{i=1}^{l}f_{n_{i}}d\mu _{n_{i}}\,,  \label{TLpsigfTs}
\end{equation}
where existence of the first and second terms is provided by adopting Eq.(%
\ref{TLpsi}) and by the rationale of NESOM-1, respectively. The higher-order
terms would be made meaningful by assuming existence of limits 
\begin{equation}
^{\left( 1\right) }\left[ P_{n_{1}}\cdot ...\cdot P_{n_{l}}\right]
_{\varepsilon }^{t}=\lim_{\left| \Omega \right| \rightarrow \infty
}\,^{\left( 1\right) }\left( P_{n_{1}}\cdot ...\cdot P_{n_{l}}\right)
_{\varepsilon }^{t}\left| \Omega \right| ^{l-1}\,,  \label{TLlthcum}
\end{equation}
so that 
\begin{equation}
\varkappa _{\varepsilon }^{\left( 1\right) }\left( t;n\right) =\lim_{\left|
\Omega \right| \rightarrow \infty }\,^{\left( 1\right) }\left\langle
P_{n}\right\rangle _{\varepsilon }^{t}  \label{TLPavneso-1}
\end{equation}
and 
\begin{equation}
\varkappa _{\varepsilon }^{\left( l\right) }\left( t;n_{1},...,n_{l}\right)
=\left( -1\right) ^{l-1}\,^{\left( 1\right) }\left[ P_{n_{1}}\cdot ...\cdot
P_{n_{l}}\right] _{\varepsilon }^{t}\,,l\geq 2.  \label{TLksupl}
\end{equation}
Note that there results a correspondence between derivative and variational
derivative with respect to $f_{n}$ in TL: 
\[
\left| \Omega \right| \dfrac{\partial }{\partial f_{n}}\rightarrow \dfrac{%
\delta }{\delta f_{n}},\;\left| \Omega \right| \rightarrow \infty , 
\]
similar to that for $F_{n}\left( t\right) $. Eq.(\ref{TLlthcum}) means,
roughly speaking, that the $l$-th order cumulant correlator for the gross
variables falls off as $\left| \Omega \right| ^{-l+1}$ at $\left| \Omega
\right| \rightarrow \infty $, which expresses the expected TL behavior of
the thermodynamic fluctuations for the intensive observables. It worth
emphasizing that adopting Eqs.(\ref{TLpsi}) and Eqs.(\ref{TLlthcum}) is by
no means sufficient for existence of $\psi _{\varepsilon }\left( t;{\frak f}%
\right) $ since it does not guarantee the convergence of the functional
Taylor series in Eq.(\ref{TLpsigfTs})

Finally, consider $\varepsilon $L of $\psi _{\varepsilon }\left( t;{\frak f}%
\right) $%
\begin{equation}
\psi \left( t;{\frak f}\right) =\lim_{\varepsilon \downarrow 0}\psi
_{\varepsilon }\left( t;{\frak f}\right) .  \label{eLpsigf}
\end{equation}
Taking formally $\varepsilon $L in each term of the series in Eq.(\ref
{TLpsigfTs}) with adopting Eq.(\ref{TLpsi}) and using the limit 
\begin{equation}
\varkappa ^{\left( 1\right) }\left( t;n_{1}\right) =\lim_{\varepsilon
\downarrow 0}\varkappa _{\varepsilon }^{\left( 1\right) }\left(
t;n_{1}\right) =\,^{\left( 1\right) }\left\langle P_{n_{1}}\right\rangle ^{t}
\label{eLPavneso-1}
\end{equation}
that exists in NESOM-1 unconditionally, gives 
\begin{equation}
\psi \left( t;{\frak f}\right) =\psi \left( t\right) +\sum_{l=1}^{\infty }%
\frac{1}{l!}\int ...\int \varkappa ^{\left( l\right) }\left(
t;n_{1},...,n_{l}\right) \prod_{i=1}^{l}f_{n_{i}}d\mu _{n_{i}}\,,
\label{eLpsigfTs}
\end{equation}
provided that the quotients of terms with $l\geq 2$%
\begin{eqnarray}
\varkappa ^{\left( l\right) }\left( t;n_{1},...,n_{l}\right)
&=&\lim_{\varepsilon \downarrow 0}\,\varkappa _{\varepsilon }^{\left(
l\right) }\left( t;n_{1},...,n_{l}\right) =\left( -1\right) ^{l-1}\,^{\left(
1\right) }\left[ P_{n_{1}}\cdot ...\cdot P_{n_{l}}\right] ^{t},
\label{eLksupl} \\
\left[ P_{n_{1}}\cdot ...\cdot P_{n_{l}}\right] ^{t} &=&\lim_{\varepsilon
\downarrow 0}\,^{(1)}\left[ P_{n_{1}}\cdot ...\cdot P_{n_{l}}\right]
_{\varepsilon }^{t}  \label{eLlthcum}
\end{eqnarray}
at least exist. Again, this condition is not sufficient for $\psi \left( t;%
{\frak f}\right) $ to exist. To proceed with the proof at goal, the
following ultimate statement is postulated.

\begin{axiom}
$\psi \left( t;{\frak f}\right) $ exists for a set of macro-parameters {\em (%
}including the both BE solutions at least{\em )} at non-trivial ${\frak f}$%
's in a vicinity of the point ${\frak f}$ $=0$ and is twice functionally
differentiable at that point.
\end{axiom}

\begin{remark}
{\em Thus, the functional }$\psi \left( t;{\frak f}\right) ${\em \ is not
required to be analytic at }${\frak f}${\em \ }$=0${\em . It follows with
necessity from Axiom 4 and Eqs.(\ref{eLPavneso-1})-(\ref{eLlthcum}) that} 
\begin{eqnarray}
\left. \frac{\delta \psi \left( t;{\frak f}\right) }{\delta f_{n}}\right| _{%
{\frak f}=0} &=&\,^{\left( 1\right) }\left\langle P_{n}\right\rangle ^{t}
\label{1stvdeLpsi} \\
\left. \frac{\delta ^{2}\psi \left( t;{\frak f}\right) }{\delta f_{n}\delta
f_{m}}\right| _{{\frak f}=0} &=&\,-\,^{\left( 1\right) }\left[ P_{n}\cdot
P_{m}\right] ^{t},  \label{2ndvdeLpsi}
\end{eqnarray}
{\em but existence of the higher correlators is not necessary for the
forthcoming proof.}
\end{remark}

\subsection{Proof}

Returning now to Eq.(\ref{PBineq}), put $A=\;S_{\varepsilon }\left( t;{\frak %
f}\right) $ and $B=S\left( t_{0},t_{0}-t\right) $ with an arbitrary $t_{0}$.
Then two inequalities result. The first is 
\[
\Psi _{\varepsilon }\left( t;{\frak f}\right) \,>\sum_{n}f_{n}\frac{\partial
\Psi _{\varepsilon }\left( t;{\frak f}\right) }{\partial f_{n}}+\frac{\,%
\mbox{Tr}\left\{ e^{-S_{\varepsilon }\left( t;{\frak f}\right) }\left[ 
\stackrel{\varepsilon }{\widetilde{S\left( t,0\right) }}-\;S\left(
t_{0},t_{0}-t\right) \right] \right\} }{\mbox{Tr}\left[ e^{-S_{\varepsilon
}\left( t;{\frak f}\right) }\right] } 
\]
Multiplication this inequality by $\varepsilon e^{\varepsilon \left(
t_{0}-t\right) }>0$ with subsequent integration over $t_{0}$ from $-\infty $
to $t$ gives 
\begin{equation}
\Psi _{\varepsilon }\left( t;{\frak f}\right) \,>\sum_{n}\frac{\partial \Psi
_{\varepsilon }\left( t;{\frak f}\right) }{\partial f_{n}}f_{n}.
\label{lowbd}
\end{equation}
The second inequality is 
\[
\Psi _{\varepsilon }\left( t;{\frak f}\right) \,<\sum_{n}f_{n}\mbox{Tr}%
\left[ e^{-S\left( t_{0},t_{0}-t\right) }P_{n}\right] +\mbox{Tr}\left[
e^{-S\left( t_{0},t_{0}-t\right) }\stackrel{\varepsilon }{\widetilde{S\left(
t,0\right) }\ }\right] -\left\langle S\left( t_{0},0\right) \right\rangle
_{q}^{t_{0}} 
\]
$\,$Multiplication of this inequality by $\eta e^{\eta \left( t_{0}-t\right)
}$, where $\eta >0$ and is arbitrary in any other respect, with subsequent
integration over $t_{0}$ from $-\infty $ to $t$ gives 
\[
\Psi _{\varepsilon }\left( t;{\frak f}\right) \,-\sum_{n}\,^{\left( 2\right)
}\!\left\langle P_{n}\right\rangle _{\eta }^{t}f_{n}<\varepsilon
\int\limits_{-\infty }^{t}\,^{\left( 2\right) }\!\left\langle S\left(
t_{0},0\right) \right\rangle _{\eta }^{t_{0}}e^{\varepsilon \left(
t_{0}-t\right) }dt_{0}\,-\,\stackrel{\eta }{\widetilde{\Sigma \left(
t\right) }} 
\]
\begin{equation}
+\,\varepsilon \int\limits_{-\infty }^{t}\sum_{n}F_{n}\left( t_{0}\right)
\,\left[ ^{\left( 2\right) }\!\left\langle P_{n}\left( t_{0},t_{0}-t\right)
\right\rangle _{\eta }^{t}-\,^{\left( 2\right) }\!\left\langle
P_{n}\right\rangle _{\eta }^{t_{0}}\right] e^{\varepsilon \left(
t_{0}-t\right) }dt_{0}.  \label{upbd}
\end{equation}
From this point forward, the notation of quasi-invariant is applied to usual
functions of time, for brevity.

Dividing by $\left| \Omega \right| $, perform TL in Eqs.(\ref{lowbd}) and (%
\ref{upbd}) with the use of Axioms 1-3, Eqs.(\ref{spqeentr}) - (\ref
{spneentr}) and the above stated correspondence between derivatives. This
results in 
\begin{equation}
\psi _{\varepsilon }\left( t;{\frak f}\right) \,\geq \int \frac{\delta \psi
_{\varepsilon }\left( t;{\frak f}\right) }{\delta f_{n}}f_{n}d\mu _{n},
\label{Llowbd}
\end{equation}
which does reconfirm non-negativity of $\psi _{\varepsilon }\left( t\right)
=\psi _{\varepsilon }\left( t;0\right) $ obtained above, and in 
\[
\psi _{\varepsilon }\left( t;{\frak f}\right) \,\leq \int \,^{\left(
2\right) }\!\left\langle P_{n}\right\rangle _{\eta }^{t}f_{n}d\mu _{n}+%
\stackrel{\varepsilon }{\,\widetilde{^{\left( 2\right) }\sigma _{\eta
}\left( t\right) }}-\stackrel{\eta }{\widetilde{\sigma \left( t\right) }} 
\]
\[
+\,\varepsilon \int\limits_{-\infty }^{t}e^{\varepsilon \left(
t_{0}-t\right) }\left\{ \int F_{n}\left( t_{0}\right) \,\lim_{\left| \Omega
\right| \rightarrow \infty }\left[ \,^{\left( 2\right) }\!\left\langle
P_{n}\left( t_{0},t_{0}-t\right) \right\rangle _{\eta }^{t}-\,^{\left(
2\right) }\!\left\langle P_{n}\right\rangle ^{t_{0}}\right] d\mu
_{n}\right\} dt_{0}. 
\]
When keeping in the latter inequality $\varepsilon $ finite, $\eta $ may
freely be tended to zero in its rhs without affecting the lhs which does not
depend on $\eta $ at all. Let $\eta $ to take values of any infinitesimal
sequence, for which the sequence $\stackrel{\eta }{\widetilde{\sigma \left(
t\right) }}$ converge to its upper limit. This limit process, on account of
Eq.(\ref{Lspneentr}) and Eq.(\ref{propneso-2}), leads to the interim bound 
\begin{equation}
\psi _{\varepsilon }\left( t;{\frak f}\right) \,\leq \int \,^{\left(
2\right) }\!\left\langle P_{n}\right\rangle ^{t}f_{n}d\mu _{n}\,+\,\stackrel{%
\varepsilon }{\widetilde{^{\left( 2\right) }\sigma \left( t\right) }}-\,\,%
\overline{\lim_{\eta \downarrow 0}}\!\stackrel{\eta }{\widetilde{\sigma
\left( t\right) }}.  \label{interim3}
\end{equation}
Let $\varepsilon $ takes values of any infinitesimal sequence, for which the
sequence $\stackrel{\varepsilon }{\widetilde{^{\left( 2\right) }\sigma
\left( t\right) }}$ converges to its upper limit. Then Eq.(\ref{interim3})
results in 
\begin{equation}
\overline{\psi }\left( t;{\frak f}\right) =\,\,\,\overline{\lim_{\varepsilon
\downarrow 0}}\,\psi _{\varepsilon }\left( t;{\frak f}\right) \,\leq \int
\,^{\left( 2\right) }\left\langle P_{n}\right\rangle ^{t}f_{n}d\mu
_{n}\,+\,\,\,\,\overline{\lim_{\varepsilon \downarrow 0}}\,\stackrel{%
\varepsilon }{\widetilde{^{\left( 2\right) }\sigma \left( t\right) }}%
-\,\,\,\,\overline{\lim_{\varepsilon \downarrow 0}}\!\stackrel{\varepsilon }{%
\widetilde{\sigma \left( t\right) }}  \label{interim4}
\end{equation}

Let now $F_{n}\left( t_{0}\right) =\,^{\left( 2\right) }F_{n}\left(
t_{0}\right) $. Then, in the rhs of Eq(\ref{interim4}) $^{\left( 2\right)
}\!\left\langle P_{n}\right\rangle ^{t}=\left\langle P_{n}\right\rangle ^{t}$%
, while the second and third terms cancel each other to give 
\begin{equation}
\,\overline{\psi }\left( t;{\frak f}\right) \leq \int \,\left\langle
P_{n}\right\rangle ^{t}f_{n}d\mu _{n}\,  \label{interim5}
\end{equation}
At ${\frak f}=0$ Eq.(\ref{interim5}) shows that 
\[
\,\overline{\lim_{\varepsilon \downarrow 0}}\,\psi _{\varepsilon }\left(
t\right) =\overline{\psi }\left( t;0\right) \leq 0, 
\]
while this limit should be non negative due to $\psi _{\varepsilon }\left(
t\right) \geq 0$. It is thus proved that on the BE solutions of NESOM-2 $\,%
\overline{\lim\limits_{\varepsilon \downarrow 0}}\,\psi _{\varepsilon
}\left( t\right) =0$. So in this case{\em \ }$\psi $ {\em does exist and
equals zero}. It is quite remarkable as existence of $\,\psi \left( t;{\frak %
f}\right) $ has not yet been invoked, but for finishing the proof it cannot
be avoided. When adopting Axiom 4, Eq.(\ref{interim5}) acquires the form 
\begin{equation}
\chi \left( t;{\frak f}\right) =\,\,\psi \left( t;{\frak f}\right) \,-\int
\,\left\langle P_{n}\right\rangle ^{t}f_{n}d\mu _{n}\leq \,\,0.
\label{Lupbd}
\end{equation}
and $\chi \left( t;0\right) =\,\psi =0$. This means that, while BE of
NESOM-2 are satisfied, the functional $\chi \left( t;{\frak f}\right) $ {\em %
becomes non-positive} {\em attaining at} ${\frak f}=0$ {\em absolute maximum 
}({\em equal zero}). Then using the necessary condition of functional
extremum and Eq.(\ref{1stvdeLpsi}) gives 
\begin{equation}
0=\left. \frac{\delta \chi \left( t;{\frak f}\right) }{\delta f_{n}}\right|
_{{\frak f}=0}=\,\left. \frac{\delta \psi \left( t;{\frak f}\right) }{\delta
f_{n}}\right| _{{\frak f}=0}-\left\langle P_{n}\right\rangle ^{t}=\,^{\left(
1\right) }\left\langle P_{n}\right\rangle ^{t}-\left\langle
P_{n}\right\rangle ^{t},  \label{extrcond}
\end{equation}
that is BE of NESOM-1. Thus every BE solution of NESOM-2 proves to satisfy
also BE of NESOM-1, i.e. $\,^{\left( 1\right) }F_{n}\left( t\right)
=\,^{\left( 2\right) }F_{n}\left( t\right) $, but {\em not vice versa}. That
the above extremum is indeed {\em maximum }is guaranteed by the second
variation of $\chi \left( t;{\frak f}\right) $ at the extremum, i.e. 
\[
\,\delta ^{\left( 2\right) }\chi \left( t;\Delta {\frak f}\right) =\int \int
\left. \frac{\delta ^{2}\chi ^{*}\left( t;{\frak f}\right) }{\delta
f_{n}\delta f_{m}}\right| _{{\frak f}=0}\delta f_{n}\delta f_{m}d\mu
_{n}d\mu _{m}=-\,\,^{(1)}\left[ {\cal P}\left( \delta {\frak f}\right) \cdot 
{\cal P}\left( \delta {\frak f}\right) \right] ^{t}\,<0, 
\]
where Eq.(\ref{2ndvdeLpsi}) was used, and ${\cal P}\left( \delta {\frak f}%
\right) =\int P_{n}\delta f_{n}d\mu _{n}$. The negativity of $\delta
^{\left( 2\right) }\chi \left( t;\delta {\frak f}\right) $ holds due to
unconditional positivity of the Kubo-Duhamel auto-correlator \cite{ruelle1}, 
\cite[Preface]{ruelle2}. This proves the equivalence of NESOM-1 and NESOM-2,
if the BE solutions in both methods are unique.%
%TCIMACRO{\TeXButton{End Proof}{\endproof}}
%BeginExpansion
\endproof%
%EndExpansion

The equivalence in the sense of Eq.(\ref{maxequiv}) is also proven using the
techique developed above, provided that $A\in \left\{ A_{p}\right\} $, where 
$A_{p}$ are some intensive variables (other than gross ones), which index $p$
satisfies Axiom 1, and the functional 
\[
\psi _{A}\left( t;{\frak g}\right) =-\lim_{\varepsilon \downarrow
0}\lim_{\left| \Omega \right| \rightarrow \infty }\frac{1}{\left| \Omega
\right| }\ln \mbox{Tr}\left[ e^{-\stackrel{\varepsilon }{\widetilde{S\left(
t,0\right) \ }}-\sum\limits_{p}g_{p}A_{p}\,}\right] 
\]
on a set ${\frak g}=\left\{ g_{p}\right\} $ may be constructed to satisfy
Axiom 4. In the present framework more definite description of variables,
for which Eq.(\ref{maxequiv}) holds, seems hardly possible.

\section{Conclusion}

To conclude, the non-pertutbative proof of equivalence between two
non-equilibrium ensembles in NESOM, based on MaxEnt principle, is proposed.
The proof is thought as an improvement upon previous ones \cite{kalashnikov2}%
, \cite{tischenko1} and \cite[Preface]{bitensky}. Because the rationales of
NESOM was not clearly delineated, some natural assumptions concerning TL
and, inherent to NESOM, $\varepsilon $L are introduced. The present proof,
is not also free of some `extra' assumption, namely existence and
second-order differentiability of the generating functional $\psi \left( t;%
{\frak f}\right) $ However, the latter seems natural in the
field-theoretical context and much more appropriate than the `time
correlation weackening' \cite{kalashnikov2}, \cite{tischenko1},
entropy-production series convergence \cite{tischenko1} and `asymptotic
trace normalisation' \cite[Preface]{bitensky} conditions, which conflict
with the basics of NESOM.

\bigskip

\noindent \noindent {\bf {\LARGE Appendix I}}

Consider `pre-limit' differential form of BE in NESOM-1. Multiplying both
sides of Eq.(\ref{eqneso-1}) by $P_{n}$ and taking the trace gives 
\begin{equation}
\frac{\partial }{\partial t}\,\left\langle P_{n}\right\rangle
_{q}^{t}-{}^{(1)}\left\langle \stackrel{\bullet }{P_{n}}\left( t,0\right)
\right\rangle _{\varepsilon }^{t}\,={\frak J}_{n,\varepsilon }\left(
t\right) ,  \label{baleqfc}
\end{equation}
where 
\begin{eqnarray}
{\frak J}_{n,\varepsilon }\left( t\right) &=&\varepsilon \,\left[ ^{\left(
1\right) }\left( P_{n}\cdot S\left( t,0\right) \right) _{\varepsilon
}^{t}\,-\,^{\left( 1\right) }\left( P_{n}\cdot \,\stackrel{\varepsilon }{%
\widetilde{S\left( t,0\right) \ }}\right) _{\varepsilon }^{t}\right] 
\nonumber \\
&=&\varepsilon \int\limits_{-\infty }^{t}\,^{\left( 1\right) }\left(
P_{n}\cdot \,\stackrel{\bullet }{S}\left( t_{0},t_{0}-t\right) \right)
_{\varepsilon }^{t}\,e^{\varepsilon \left( t_{0}-t\right) }dt_{0}.
\label{befsource}
\end{eqnarray}
Thus, for Eq.(\ref{difbaleqf}) to hold at $\alpha =1$, it is necessary that $%
{\frak J}_{n,\varepsilon }\left( t\right) =0$. However, any connection of
the resulting from this projective-type equations with Eq.(\ref{baleqf}) can
hardly be stated. Note that weaker condition 
\begin{equation}
\lim_{\varepsilon \downarrow 0}\,\lim_{\left| \Omega \right| \rightarrow
\infty }{\frak J}_{n,\varepsilon }\left( t\right) =0  \label{corrweak}
\end{equation}
appears in Ref.\cite{kalashnikov2} as a condition for the equivalence of
NESOM-1 and NESOM-2. With Eq.(\ref{corrweak}), Eq.(\ref{difbaleq}) for $%
\alpha =1$ holds trivially, but connection of Eq.(\ref{baleq}) and that
condition remains obscure. The key problem here, as seen from Eq.(\ref
{befsource}), is the time behaviour of either $^{\left( 1\right) }\left(
P_{n}\cdot \,P_{m}\left( t_{0},t_{0}-t\right) \right) _{\varepsilon }^{t}$
or $^{\left( 1\right) }\left( P_{n}\cdot \,\stackrel{\bullet }{S}\left(
t_{0},t_{0}-t\right) \right) _{\varepsilon }^{t}$, which should be at least
bounded or fall off at $t_{0}\rightarrow -\infty $, respectively. These
conditions are not found among the rationales of NESOM.

\bigskip

\noindent {\bf {\LARGE Appendix II}}\noindent

This Appendix overviews previous treatments of the equivalence problem.

Consider first the proof of Ref.\cite{kalashnikov2}. The `time
correlation-weakening' condition of Ref.\cite{kalashnikov2} is essentially
the same as Eq.(\ref{corrweak}) and implies that $\lim\limits_{\varepsilon
\downarrow 0}\,\lim\limits_{\left| \Omega \right| \rightarrow \infty
}\,^{\left( 1\right) }\left( P_{n}\cdot \,\stackrel{\bullet }{S}\left(
t_{0},t_{0}-t\right) \right) _{\varepsilon }^{t}\rightarrow 0$ as $%
t_{0}\rightarrow -\infty $. However, the NESOM perturbational practice
evidences that such a behavior shows up only {\em after} BE were used to
eliminate the terms $\propto \,\stackrel{\bullet }{F}_{m}\left( t_{0}\right) 
$ off $\stackrel{\bullet }{S}\left( t_{0},t_{0}-t\right) $. Thus, the
condition of Ref.\cite{kalashnikov2} may be {\em necessary}, {\em but by no
means sufficient}, for prooving the equivalence.%
%TCIMACRO{\TeXButton{End Proof}{\endproof}}
%BeginExpansion
\endproof%
%EndExpansion

Consider next the proof of Ref.\cite{tischenko1} (also \cite[Preface]
{tischenko2}) claimed for classical case. Here the approach of Ref.\cite
{tischenko1} is extended to quantum case. Let us obtain formal expansion of $%
^{\left( \alpha \right) }\rho _{\varepsilon }\left( t\right) $ with respect
the entropy production. Making use of the last equality in Eq.(\ref{quasinv}%
), expand the operator exponent defining QNESO-1 and its trace to obtain 
\begin{eqnarray}
^{\left( 1\right) }\rho _{\varepsilon }\left( t\right) &=&\rho _{q}\left(
t\right) \left\{ 1+\right.  \label{interim6} \\
&&\left. \sum_{k=1}^{\infty }\frac{1}{k!}\int\limits_{-\infty
}^{0}...\int\limits_{-\infty
}^{0}\int\limits_{0}^{1}...\int\limits_{0}^{1}\left[ V_{t}\left( \lambda
_{1},\tau _{1}\right) ...\cdot V_{t}\left( \lambda _{k},\tau _{k}\right)
\right] _{\text{irr}}^{>}\prod_{j=1}^{k}d\lambda _{j}e^{\varepsilon \tau
_{j}}d\tau _{j}\right\} ,  \nonumber
\end{eqnarray}
where $V_{t}\left( \lambda ,\tau \right) \triangleq e^{\lambda S\left(
t,0\right) }\stackrel{\bullet }{S}\left( t+\tau ,\tau \right) e^{-\lambda
S\left( t,0\right) }$. Using the identity: $S\left( t+\tau _{0},\tau
_{0}\right) =S\left( t,0\right) -\int\limits_{\tau _{0}}^{0}\stackrel{%
\bullet }{S}\left( t+\tau _{1},\tau _{1}\right) d\tau _{1}$, analogous
expansion for QNESO-2 is obtained 
\begin{eqnarray}
^{\left( 2\right) }\rho _{\varepsilon }\left( t\right) &=&\rho _{q}\left(
t\right) \left\{ 1+\right.  \label{interim7} \\
&&\left. \sum_{k=1}^{\infty }\frac{1}{k!}\varepsilon \int\limits_{-\infty
}^{0}e^{\varepsilon \tau _{0}}d\tau _{0}\int\limits_{\tau
_{0}}^{0}...\int\limits_{\tau
_{0}}^{0}\int\limits_{0}^{1}...\int\limits_{0}^{1}\left[ V_{t}\left( \lambda
_{1},\tau _{1}\right) ...\cdot V_{t}\left( \lambda _{k},\tau _{k}\right)
\right] _{\text{irr}}^{>}\prod_{j=1}^{k}d\lambda _{j}d\tau _{j}\right\} . 
\nonumber
\end{eqnarray}
As $\int\limits_{0}^{1}...\int\limits_{0}^{1}\left[ V_{t}\left( \lambda
_{1},\tau _{1}\right) ...\cdot V_{t}\left( \lambda _{k},\tau _{k}\right)
\right] _{\text{irr}}^{>}\prod_{j=1}^{k}d\lambda _{j}$ is symmetric in
variables $\tau _{1},...,\tau _{k}$, the integration over them in Eqs.(\ref
{interim6}) and (\ref{interim7}) may be changed to $\int\limits_{-\infty
}^{0}d\tau _{1}\int\limits_{-\infty }^{\tau _{1}}d\tau
_{2}...\int\limits_{-\infty }^{\tau _{k-1}}d\tau _{k}$ and $%
\int\limits_{\tau _{0}}^{0}d\tau _{k}\int\limits_{\tau _{k}}^{0}d\tau
_{k-1}...\int\limits_{\tau _{2}}^{0}d\tau _{1}$, respectively, to times $k!$%
. Using the latter transformation, each term of expansion in Eq.(\ref
{interim7}) can be integrated over $\tau _{0}$ by parts, after which there
remains integration over the domain $-\infty <\tau _{k}<\tau _{k-1}<...<\tau
_{1}\leq 0$. Changing this integration order to the opposite one makes the
integration domains in both Eq.(\ref{interim6}) and Eq.(\ref{interim7}) the
same. This results in the unified expansion of QNESO 
\begin{eqnarray*}
^{\left( \alpha \right) }\rho _{\varepsilon }\left( t\right) &=&\rho
_{q}\left( t\right) \left\{ 1+\right. \\
&&\left. \sum_{k=1}^{\infty }\int\limits_{-\infty }^{0}\int\limits_{-\infty
}^{\tau _{1}}...\int\limits_{-\infty }^{\tau _{k-1}}\,^{\left( \alpha
\right) }g_{k}\int\limits_{0}^{1}...\int\limits_{0}^{1}\left[ V_{t}\left(
\lambda _{1},\tau _{1}\right) ...\cdot V_{t}\left( \lambda _{k},\tau
_{k}\right) \right] _{\text{irr}}^{>}\prod_{j=1}^{k}d\lambda _{j}d\tau
_{j}\right\} ,
\end{eqnarray*}
from which the pre $\varepsilon $L quasi-averages expansion follows 
\begin{eqnarray}
^{(\alpha )}\!\left\langle A\right\rangle _{\varepsilon }^{t}
&=&\!\left\langle A\right\rangle ^{t}+  \label{unqav} \\
&&\sum_{k=1}^{\infty }\int\limits_{-\infty }^{0}\int\limits_{-\infty }^{\tau
_{1}}...\int\limits_{-\infty }^{\tau _{k-1}}\,^{\left( \alpha \right)
}g_{k}\int\limits_{0}^{1}...\int\limits_{0}^{1}\left\langle \left[
V_{t}\left( \lambda _{1},\tau _{1}\right) ...\cdot V_{t}\left( \lambda
_{k},\tau _{k}\right) \right] _{\text{irr}}^{>}A\right\rangle
^{t}\prod_{j=1}^{k}d\lambda _{j}d\tau _{j},  \nonumber
\end{eqnarray}
where $^{\left( 1\right) }g_{k}=e^{\varepsilon \left( \tau _{1}+...+\tau
_{k}\right) }$,$\;^{\left( 2\right) }g_{k}=e^{\varepsilon \tau _{k}}$. As $%
\lim\limits_{\varepsilon \downarrow 0}\,^{\left( \alpha \right) }g_{k}=1$,
it was concluded in Ref.\cite{tischenko1} (also \cite[Preface]{tischenko2})
that Eq.(\ref{maxequiv}) holds unconditionally. In fact, this statement is
valid only if the irreducible correlators $\left\langle \left[ V_{t}\left(
\lambda _{1},\tau _{1}\right) ...\cdot V_{t}\left( \lambda _{k},\tau
_{k}\right) \right] _{\text{irr}}^{>}A\right\rangle ^{t}$ are {\em %
absolutely integable} in the domain $0>\tau _{1}>...>\tau _{k}>-\infty $,
otherwise counterexamples show that the above integrals with $\alpha =1$ and 
$\alpha =2$ many have quite different values. This is another form of the
`time correlation-weakening' condition. Again, referring to the practice of
NESOM shows that the irreducible correlators, though satisfying a `spatial
correlation-weakening' at all $F_{n}\left( t_{0}\right) $, don't satisfy
`time correlation-weakening' unless the above-mentioned exclusion of
`secular' terms is made with the use of BE. Thus, without the assumption
that BE are satisfied in advance, the proof of Ref.\cite{tischenko1} is
deceptive, while that assumption makes the proof unclosed.%
%TCIMACRO{\TeXButton{End Proof}{\endproof}}
%BeginExpansion
\endproof%
%EndExpansion

At last, consider proof of Ref.\cite[Preface]{bitensky}. The use is made of
Jensen convexity inequality that gives 
\[
\,\stackrel{\varepsilon }{\widetilde{e^{-S(t,0)}}}\,>\,e^{-\stackrel{%
\varepsilon }{\widetilde{S\left( t,0\right) \ }}} 
\]
at each phase-space point. This inequality is consistent with $\Psi
_{\varepsilon }\left( t\right) >0$, being a particular case of Eq.(\ref
{lowbd}) for ${\frak f}=0$. As a result, in classical case the following
pointwise inequality between QNESO-2 and QNESO-1 is stated 
\begin{equation}
^{\left( 2\right) }\!\rho _{\varepsilon }\left( t\right) -\,^{\left(
1\right) }\!\rho _{\varepsilon }\left( t\right) e^{-\Psi _{\varepsilon
}\left( t\right) }>0.\,  \label{nso2gnso1}
\end{equation}
Using the identity 
\[
{}^{(2)}\!\left\langle A\right\rangle _{\varepsilon
}^{t}-\,^{(1)}\!\left\langle A\right\rangle _{\varepsilon }^{t}\equiv \int
\left[ ^{\left( 2\right) }\rho _{\varepsilon }\left( t\right) -\,\,^{\left(
1\right) }\rho _{\varepsilon }\left( t\right) e^{-\Psi _{\varepsilon }\left(
t\right) }\right] Ad\omega +\left[ e^{-\Psi _{\varepsilon }\left( t\right)
}-1\right] \,^{(1)}\!\left\langle A\right\rangle _{\varepsilon }^{t}\,, 
\]
where $d\omega $ is the phase-space measure, and assuming the observable $A$
to be bounded, one obtains the estimation 
\begin{eqnarray*}
\left| ^{(2)}\!\left\langle A\right\rangle _{\varepsilon
}^{t}-\,^{(1)}\!\left\langle A\right\rangle _{\varepsilon }^{t}\right| &\leq
&\int \left| ^{\left( 2\right) }\rho _{\varepsilon }\left( t\right)
-\,\,^{\left( 1\right) }\rho _{\varepsilon }\left( t\right) e^{-\Psi
_{\varepsilon }\left( t\right) }\right| \left| A\right| d\omega +\left[
1-e^{-\Psi _{\varepsilon }\left( t\right) }\right] \,\left|
^{(1)}\!\left\langle A\right\rangle _{\varepsilon }^{t}\right| \\
&\leq &\left\| A\right\| _{\Omega }\left\{ \int \left| ^{\left( 2\right)
}\rho _{\varepsilon }\left( t\right) -\,\,^{\left( 1\right) }\rho
_{\varepsilon }\left( t\right) e^{-\Psi _{\varepsilon }\left( t\right)
}\right| d\omega +\left[ 1-e^{-\Psi _{\varepsilon }\left( t\right) }\right]
\right\} ,
\end{eqnarray*}
where $\left\| A\right\| _{\Omega }=\max \left| A\right| $, and $\Psi
_{\varepsilon }\left( t\right) >0$ is taken into account. Next, the use of
Eq.(\ref{nso2gnso1}) and the trace normalisation of QNESO gives 
\begin{equation}
\left| ^{(2)}\!\left\langle A\right\rangle _{\varepsilon
}^{t}-\,^{(1)}\!\left\langle A\right\rangle _{\varepsilon }^{t}\right| \leq
2\left\| A\right\| _{\Omega }\left[ 1-e^{-\Psi _{\varepsilon }\left(
t\right) }\right]  \label{interim8}
\end{equation}
Note that this consideration cannot be extended to quantum system, since
operator exponent is not convex operator function \cite{bendat}. Proceeding
with the above estimate, Bitensky \cite[Preface]{bitensky} assumed two
conditions: (i) independence of $\left\| A\right\| _{\Omega }$ on $\Omega $;
(ii) the `asymptotic normalisation condition' 
\begin{equation}
\lim_{\varepsilon \downarrow 0}\lim_{\left| \Omega \right| \rightarrow
\infty }\Psi _{\varepsilon }\left( t\right) =0,\text{ or }\lim\Sb %
\varepsilon \downarrow 0  \\ \left| \Omega \right| \rightarrow \infty 
\endSb \Psi _{\varepsilon }\left( t\right) =0.  \label{asymnorm}
\end{equation}
As seen from Eq.(\ref{interim8}), (i) and (ii) lead to the equivalence of
the NESOM ensembles in the sense of Eq.(\ref{maxequiv}). Although (i) is by
no means a restrictive condition, (ii) is in doubt. The latter condition is
obviously invalid at general $F_{n}\left( t_{0}\right) $, because TL and $%
\varepsilon $L of $\Psi _{\varepsilon }\left( t\right) /\left| \Omega
\right| $ (performed either sequentially or simultaneously) is strictly
positive, and so $\Psi _{\varepsilon }\left( t\right) \rightarrow \infty $
unless BE are satisfied. For TL and $\varepsilon $L perfomed sequentially
and $F_{n}\left( t_{0}\right) =\,^{\left( 1\right) }F_{n}\left( t_{0}\right) 
$, $\Psi _{\varepsilon }\left( t\right) \rightarrow \infty $, since $\psi
_{\varepsilon }\left( t\right) $ remains positive at finite $\varepsilon $.
Simaltaneous TL and $\varepsilon $L of $\Psi _{\varepsilon }\left( t\right) $
is uncertain, since that for $\Psi _{\varepsilon }\left( t\right) /\left|
\Omega \right| $ is zero. However, for non-stationary process Eq.(\ref
{asymnorm}) also proves invalid. Indeed, assume that $^{\left( 1\right) }%
{\frak s}$ exists. Then, given small $\delta >0$, large $\Delta >0$ and
arbitrary $\tau >0$, there exist $\Omega _{0}$ and $\varepsilon _{0}$
satisfying $\left| \Omega _{0}\right| \varepsilon _{0}>\Delta $ (see
footnote 1), such that 
\[
^{\left( 1\right) }\!\left\langle S\left( t_{0},0\right) \right\rangle
_{\varepsilon }^{t}-\,^{\left( 1\right) }{\frak S}_{\varepsilon }\left(
t_{0}\right) >\left| \Omega \right| \left[ \sigma \left( t_{0}\right)
-\,^{\left( 1\right) }{\frak s}-\delta \right] >0 
\]
for every $\Omega \supset \Omega _{0}$, $\varepsilon <\varepsilon _{0}$
satisfying $\left| \Omega \right| \varepsilon >\Delta $, and $t_{0}\in
\left[ t-\tau ,t\right] $. Making use of Eq.(\ref{Psi-qGentr}) and the above
inequality gives 
\[
\Psi _{\varepsilon }\left( t\right) >\Delta \tau \left[ \min_{t_{0}\in
\left[ t-\tau ,t\right] }\sigma \left( t_{0}\right) -\,^{\left( 1\right) }%
{\frak s}-\delta \right] , 
\]
so that $\Psi _{\varepsilon }\left( t\right) \rightarrow \infty $ as well.
For a stationary state, this reasoning is useless, since $\sigma =\,^{\left(
1\right) }{\frak s}$ at $F_{n}=\,^{\left( 1\right) }F_{n}$. In this case
simultaneous TL and $\varepsilon $L of $\Psi _{\varepsilon }$ remains
uncertain, but it cannot be arbitrarily fixed at zero value.%
%TCIMACRO{\TeXButton{End Proof}{\endproof}}
%BeginExpansion
\endproof%
%EndExpansion

\noindent \noindent

\noindent \baselineskip 18pt{}

\end{document}